\magnification=1200
\def\singlespace{\baselineskip 12 pt}

\def\oneandahalfspace{\baselineskip 18pt}
\def\blankline{\vskip 12 pt\noindent}

\def\secto#1\endsecto{\vskip 20pt {\bf #1}\vskip 7pt\nobreak}
\global\newcount\refno \global\refno=1
\newwrite\rfile
\def\ref#1#2{\hbox{[\hskip 2pt\the\refno\hskip 2pt]}\nref#1{#2}}
\def\nref#1#2{\xdef#1{\hbox{[\hskip 2pt\the\refno\hskip 2pt]}}%
\ifnum\refno=1\immediate\openout\rfile=refs.tmp\fi%
\immediate\write\rfile{\noexpand\item{\noexpand#1\ }#2.}%
\global\advance\refno by1}
\def\semi{;\hfil\noexpand\break}
\def\demi{:\hfil\noexpand\break}

\newwrite\efile \let\firsteqn=T
\def\writeqno#1%
{\if T\firsteqn \immediate\openout\efile=eqns.tmp\global\let\firsteqn=F\fi%
\immediate\write\efile{#1 \string#1}\global\advance\meqno by1}

\def\eqnn#1{\xdef #1{(\the\secno.\the\meqno)}\writeqno#1}
\def\eqna#1{\xdef #1##1{(\the\secno.\the\meqno##1)}\writeqno{#1{}}}

\def\eqn#1#2{\xdef #1{(\the\secno.\the\meqno)}%
$$#2\eqno(\the\secno.\the\meqno)$$\writeqno#1}
\global\newcount\meqno \global\meqno=1
\newwrite\efile \let\firsteqn=T
\def\writeqno#1%
{\if T\firsteqn \immediate\openout\efile=eqns.tmp\global\let\firsteqn=F\fi%
\immediate\write\efile{#1 \string#1}\global\advance\meqno by1}

\def\eqqn#1#2{\xdef #1{(\the\meqno)}%
$$#2\eqno(\the\meqno)$$\writeqno#1}
\def\frac#1#2{{#1 \over #2}} 
\def\pr{{Phys. Rev.}\ }
\def\nonumfirst{\nopagenumbers
                \footline={\ifnum\count0=1\hfill
                           \else\hfill\folio\hfill
                           \fi}}
\nonumfirst
 at 12truept 
\font\magoner=cmr10 at 12truept 
\font\magthreer=cmr10 at 17.28truept 
\font\magonei=cmti10 at 12truept 
\font\maghalfr=cmr10 at 10.95truept 
 at 17.28truept 
\global\newcount\meqno \global\meqno=1
\newwrite\efile \let\firsteqn=T
\def\writeqno#1%
{\if T\firsteqn \immediate\openout\efile=eqns.tmp\global\let\firsteqn=F\fi%
\immediate\write\efile{#1 \string#1}\global\advance\meqno by1}

\def\eqqn#1#2{\xdef #1{(\the\meqno)}%
$$#2\eqno(\the\meqno)$$\writeqno#1}
\vsize=25 truecm
\hsize=16 truecm
\voffset=-0.8 truecm
%
\singlespace
\parskip 6truept
\parindent 20truept
\vbox{{\rightline{\magoner UNIBAS-MATH 10/96}}
}
\hyphenation{ex-pe-ri-men-tal}
\hyphenation{va-cu-um}
\vskip 4.4truecm
\centerline{\magthreer The WKB Approximation Without Divergences}
\vskip 2truecm
\vskip 33truept
\centerline{\magoner D. Cocolicchio$^{(1,2)}$ and M. Viggiano$^{(1)}$}
\vskip 20truept
\vbox{
\centerline{\magonei $^{1)}$Dipartimento di Matematica,
Univ. Basilicata, Potenza, Italy}
\vskip 5truept
\centerline{\magonei Via N. Sauro, 85, 85100 Potenza, Italy} }
\vskip 15truept
\vbox{
\centerline{\magonei $^{2)}$Istituto Nazionale di Fisica Nucleare,
                     Sezione di Milano, Italy}
\vskip 5truept
\centerline{\magonei Via G. Celoria 16, 20133 Milano, Italy} }
\vskip 4.8truecm
\centerline{\maghalfr ABSTRACT}
\vskip 15truept
\oneandahalfspace
\noindent
In this paper, the WKB approximation to the scattering problem
is developed without the divergences which usually appear
at the classical turning points. A detailed procedure of
complexification is shown to generate results identical to the usual
WKB prescription but without the cumbersome connection formulas.
\vskip 1truecm
\noindent
\singlespace
%
\vfill\eject
\vsize=24 truecm
\hsize=16 truecm
\baselineskip 18 truept
\parindent=1cm
\parskip=8pt
\phantom{.}
\blankline
\leftline {\bf I. Introduction}
\blankline
\noindent
In general an exact solution for many quantum mechanical problems is 
unavoidable and one is forced to resort to some type of approximate 
technique.
One of the most useful of those methods is the semiclassical or
Wentzel-Kramers-Brillouin (WKB) approach
\ref\WKB{G. Wentzel, Z. Phys. {\bf 38} (1926) 518; H. A. Kramers, Z. Phys.
{\bf 39} (1926) 828; L. Brillouin, J. Phys. (Paris) {\bf 7} (1926) 353}.
The major shortcoming of the semiclassical WKB approximation of solving
the wave equation is its divergence at the classical turning points.
Presently, available regularization schemes are accurate, but rather
complicated.
Although these methods sharpen the threshold effects, nevertheless
exact solutions to the stationary wave equation 
\eqqn\eIO{
{{d^2\psi}\over{dx^2}}+ k^2(x) \psi=0
}
(with the local wavenumber $k^2(x)={2m[E-V(x)]}/{\hbar^2}$ ) cannot be 
found in most problems which involve one-dimensional potential $V(x)$
\ref\LPL{L. P. Eisenhart, Phys. Rev. {\bf 45} (1934) 427}.
In the WKB approach, the wave function $\psi(x)$ is supposed to be 
represented by
\eqqn\eIU{
\psi(x)=Ae^{i\sigma(x)/\hbar}
}
 which converts the linear, time-independent
Schr\"odinger equation for $ \psi(x)$ into the non-linear Riccati equation
for the function $\sigma(x)$ 
\blankline
\eqqn\eID{
\left( {{d\sigma}\over{dx}}\right)^2-i\hbar { {d^2\sigma}\over
{dx^2}}=2m[E-V(x)] \quad .
}
\blankline
The WKB approximation consists in
expanding $\sigma $ as a power series in $\hbar$:
\blankline
\eqqn\eIT{ 
\sigma(x)  \simeq {\sigma_0}(x)+{{\hbar}\over{i}}{\sigma_1}(x)+
{\left({{\hbar}\over{i}}\right)}^2{\sigma_2}(x)+o(\hbar^3)
}
\blankline
and substituting this expression in the relative differential equation,
whose  coefficients can be generated recursively
\ref\JD{J. Dunham, Phys. Rev. {\bf 41} (1932) 721}.
The following conditions can then be obtained:
\eqqn\eITB{
\cases{
{\left({{d{\sigma}_0 (x)}\over{dx}}\right)}^ 2 = 2m[E-V(x)]\equiv p^ 2(x)
\cr
   \cr
{\left({{d{\sigma}_0 (x)}\over{dx}}\right)}
{\left({{d{\sigma}_1 (x)}\over{dx}}\right)}
+ {1\over 2} {\left({{d^2{\sigma}_0 (x)}\over{dx^2}}\right)} = 0\cr }
}
whose solutions are given by 
\eqqn\eIQB{
\eqalign{
{\sigma_0}(x)=&\pm \int_{x_0}^ x p(x') \, dx'\cr
{\sigma_1^\prime}(x)=& -{{1}\over{2}}{{{\sigma_0}^{\prime\prime} (x)}
\over{{\sigma_0^\prime}(x)}}=
-{{1}\over{2}}{{{p}^\prime (x)}\over{{p}(x)}}\Rightarrow
\;\; \sigma_1 (x) =-\ln\sqrt{p(x)}\cr } }
where the prime denotes differentiation with respect to $x$  and $x_0$ is 
an arbitrary point.
The leading connection term $\sigma_1(x)$ diverges at the classical
turning points $x^{(i)}_c$ where $V(x^{(i)}_c)=E$, which makes 
the first-order WKB solution
\blankline
\eqqn\eICI{
\psi(x)\simeq {{1}\over{\sqrt{p(x)}}}\left\{{C_+}\exp{\left({{i}\over{\hbar}}
 \int_{x_0}^ x p(x')\, dx'
\right)}+{C_-}\exp{\left(-{{i}\over{\hbar}}
 \int_{x_0}^ x p(x')\, dx'\right)}\right\}
}\blankline
divergent in these points.
In the classical limit, this divergence is understandable, since a 
classical particle has zero velocity at these turning points.
This divergence at the turning points is a severe limitation on the
usefulness of the WKB approximation for quantum mechanics, since there is
no divergence in the exact wave function.
However away from the turning points, the WKB solution gives a good 
description of wave functions, especially in the semiclassical limit of
large quantum numbers $n$.
Far from the turning points, the behaviour of the semiclassical solution 
\blankline
\eqqn\eIs{
\psi(x)={{1}\over{{\sqrt{p(x)}}}}
\left\{{C_+} \exp{\left(i \int_{x_0}^ x k(x') \, dx'\right)}+
{C_-} \exp{\left(-i \int_{x_0}^ x k(x') \, dx'
\right)}\right\}
}
\blankline
changes drastically in accord with the wave-number
\blankline
\eqqn\uni{
k(x)={{1}\over{\hbar}}p(x)={{1}\over{\hbar}}\sqrt{
2m[E-V(x)]} \quad . }
An oscillatory behaviour is produced by the solution corresponding to
the classically allowed region $E>V_{max}(x)$, where $ k(x)$ is real:
\blankline
\eqqn\dui
{\psi(x)=
{{A}\over{\sqrt{p(x)}}}\sin\left\{\int_{x_0}^  x k(x')\, dx' +
{{\pi}\over4} \right\} 
}
whereas in the classically forbidden region $E<V_{min}(x)$ where the 
wave-number $k(x)=i\beta(x)$ with 
\eqqn\tri{
\beta(x)={{1}\over{\hbar}}\sqrt{2m[V(x)-E]}>0 \quad , }
\blankline
becomes purely imaginary, the general semiclassical solution is 
exponentially decrescent: 
\blankline
\eqqn\qua{
\psi(x)={{1}\over{\sqrt{\beta(x)}}}\left\{{C_+} \exp{\left(-\int_{x_0}^ x
\beta(x') \, dx' \right)}+{C_-} 
\exp{\left(\int_{x_0}^ x \beta(x') \, dx' \right)}\right\} \quad . }
This semiclassical behaviour is valid only asymptotically.
The regions near the turning points should be treated separately.
This leads to the fragmentation of the $x$  axis into several regions with
connection formulas for going through the turning points.
Such patchwork for the semiclassical wave functions leads to the 
familiar lowest-order WKB energy quantization condition 
\eqqn\utni{\int_{x^{(1)}_c}^ {x^ {(2)}_ c}
 {\sqrt{2m[E_n-V(x)]}}
\, dx=\left(n+{{1}\over{2}}\right){\pi}\hbar
}
or to more complicated conditions if higher orders 
corrections in $\hbar$ are kept
\ref\KR{
D. T. Barclay, ''{\it Convergent WKB Series}'',
preprint Univ. Illinois, Chicago UIC HEP-TH/93-16 
(hep-th 9311169)}.
On the other side, continous connection formulas giving finite wave 
functions also at the turning points have also been developed in the 
realm of uniform approximation
\ref\RFCH{
M. V. Berry, K. E. Mount, Rep. Prog. Phys. {\bf 35} (1972) 
315; A. Voros, Ann. Inst. Henry Poincar\`e, {\bf A24} (1976) 31;
F. Robicheaux, U. Fano, M. Cavagnero and D. A. Harmin,
\pr {\bf A35} (1987) 3619}.
In the next section we recover the connection formulas of the usual
WKB procedure, then in a later section we establish that this prescription
can be generalized with a reformulation
based on the analytic continuation into
the complex momentum variable and contour integrations.
\blankline
\leftline {\bf II. Semiclassical Approach to Barrier Penetration}
\blankline
In the case of scattering problems,
there are two independent solutions which are
usually called incoming and outgoing waves, respectively.
These waves become free-particle waves in the asymptotic region.
Classically a particle (incident from the left with energy $E<V_{max}
(x)$) is completely reflected from the potential region at the (left-
hand) classically turning point $a$, defined by 
\eqqn\cini{
V(a)=E \quad . }
However, quantum mechanically the particle can "tunnel" through the
barrier and find itself on the right-hand side.
In the case of ordinary barrier penetration $V_{min}<E<V_{max}$ there
exist two classical turning points on the real axis $x^{(1)}_c=a<x^{(2)
}_c=b$.
The WKB approximation is valid where transmission dominates over 
reflection and, in the semiclassical limit, the probability
of tunneling is given by $T=e^{-2\sigma_*}$ where
\eqqn\uat{
\sigma_*=\int_a^ b \beta(x)\,dx=
{{1}\over{\hbar}}{\int_a^ b}\sqrt{2m[V(x)-E]}\, dx}
and the probability of being reflected is correspondingly reduced from
its classical value of unity to $R=1-T.$
The problem of the breakdown of the WKB solution near the turning points
can be overcome with the usual procedure to consider the linear 
approximation of the potential
\eqqn\aaa{
V(x) = V(a) + \mu(x-a) }
with
\eqqn\ader{
\mu = \left({dV}\over{dx}\right)_{x=a} }
and
\eqqn\jki{
p(x) = \sqrt{2m\mu(x-a)} \quad . }
Thus, the connection formulas, which relate oscillatory and exponential
behaviour of the wavefunction forms on the opposite sides of a classical
turning point, can be matched by considering the differential equation
\eqqn\aabb{
\left({d^2 \over{dz^2}} - z \right) \psi(z) = 0 }
once we set $z={\left({2m{\mu}}/{{\hbar}^2}\right)}^{1/3}(x-a).$
The complex method for treating classical turning points provides a 
powerful description to the "connection problem" by using the exact 
solution of the Schr\"odinger's equation
\blankline
\eqqn\essI{
\psi(x)= {{1}\over{\sqrt{p(x)}}}\left\{{C_+}(x)\exp{\left({{i}\over{\hbar}}
 w(x_0,x)
\right)}+{C_-}(x)\exp{\left(-{{i}\over{\hbar}}
 w(x_0,x)\right)}\right\}
}
where
\eqqn\eday{
w(x_0,x)=\int_{x_0}^ x p(x')\, dx'.}
The usual WKB method, by its very nature, cannot take into systematic
account the modifying effects of the multiply reflections between the
classical turning points.
Even if the values of the WKB multipliers $C_{\pm}(x)$ are known on a 
given wide region, the complex method, while providing useful information
about the solutions, it results inefficient to give all the details 
of the wave function in the neighbourhood of the turning points.
A powerful technique to overcome this deficiency consists in deriving 
the asymptotic solutions of the differential Airy equation (19)
calculated in the stationary phase approximation.
Its solution can be represented by the Laplace integral
\eqqn\ercx{
\psi(x)=A\int\limits_C {e^ {zt-{t^3}/3}}\, dt}
where the curve $C$ is taken so that the integrand vanishes at the limit
of integration. Since the integrand vanishes exponentially in the 
following interval
\eqqn\pjgt{
|arg{\,} t|<{{\pi}\over 6}}
\eqqn\gvbsm{
{{\pi}\over 2} < arg\, t < {5\over6}\pi}
\eqqn\hjfes{
{7\over 6}\pi < arg \,  t < {3\over 2}\pi}
the curves $C_1$, $C_2$, $C_3$ are all allowed and yield three different
solutions and two each are independent. The integration along imaginary 
axis
$C_1$ yields the so called Airy function
\blankline 
\eqqn\ugyka{
Ai(z)=
{1\over{2{\pi}i}} {\int\limits_{C_1} \exp{\left(zt-{{t^3}\over{3}}
\right)}\, dt}=
{1\over{\pi}}{\int_0^ {+\infty} \cos{\left(zt+ {{t^3}
\over{3}}\right)}\, dt }
}
whereas the integration along a complementary curve
$C_2$ yields a further independent
solution
\blankline
\eqqn\tgfwk{
Bi(z)=
{1\over{2i}} {\int\limits_{C_2} \exp{\left(zt-{{t^3}\over{3}}
\right)}\, dt}=
{\int_0^ {+\infty} \sin{\left(zt+ {{t^3}
\over{3}}\right)}\, dt }
}\blankline 
whose well-known asymptotic behaviours are given by
\eqqn\fvsl{
Ai(z) \sim {{1}\over {{2{\pi} {z^{1/4}}}}}
{\exp{\left(-{{2}\over {3}}
{z^{3/2}}\right)}} }

\eqqn\Lliy{
Ai(-z) \sim {{1}\over {{\pi} {z^{1/4}}}}
{\sin{\left(
{{2}\over {3}} {z^{3/2}}+{ {\pi}\over{4} } \right)}} \quad .}
Similarly in the case of $Bi(z)$ we have 
\eqqn\trfd{
Bi(z) \sim {1\over {z^{1/4}}}{\exp{\left({2 \over 3}
{z^{3/2}} \right)}} }
\eqqn\dlre{
Bi(-z) \sim {1\over {z^{1/4}}}{\cos{\left({2 \over 3}
{z^{3/2}} + {{\pi}\over 4} \right)}} \quad . }
These two Airy functions are independent solutions of the Eq.(19).
If we introduce the variable 
\eqqn\bess{
{\tau}={2\over3}{(-z)^{3/2}}
}
and make the transformation
\eqqn\jkdsw{
{\psi(z)}={(-z)^{1/2}}{\phi(\tau)}}
Eq.(19) takes the Bessel form
\eqqn\beseq{
{\tau}^{2}{{{d^2}\phi}\over{d{\tau}^2}}+{\tau}{{d\phi}\over{d\tau}}
+{\left({\tau^2}-{1\over9}\right)}\phi(\tau)=0}
which let us represent Airy functions in terms of the following
Bessel functions
\eqqn\iaib{
Ai(z)={{\sqrt z}\over{3}}\left\{{I_{-1/3}}\left({2\over3}z^{3/2}
\right)-{I_{1/3}}\left({2\over3}z^{3/2}\right)\right\}}

\eqqn\abii{
Bi(z)={\sqrt{z\over3}}\left\{{I_{-1/3}}\left({2\over3}z^{3/2}
\right)+{I_{1/3}}\left({2\over3}z^{3/2}\right)\right\} \quad .}
The formal connection formulas can be established by the analytic
features of the potential barriers near a classical turning point
to match approximate solutions across the boundaries. The derivation
and application of the connection formulas are both non trivial and
fraught with pitfalls associated with the existence of exponentially
large and exponentially small components of the wave function, in the 
classically forbidden region. However in spite of these difficulties,
in the case $z>0$   $({dV}/{dx}>0)$ we have the local wavenumber in
the forbidden region
\blankline
\eqqn\tghy{
\beta(x) = {1\over{\hbar}} {\sqrt{2m[V(x)-E]}} = 
{1\over{\hbar}} {\sqrt{2m{\mu}(x-a)}} }
so that
\eqqn\lkju{
\int_a^ x \beta(x')\, dx' = {2 \over 3}{z^{3/2}} \quad . }
Similarly for $z<0$ we have in the allowed region
\blankline
\eqqn\fred{
k(x) = {1\over{\hbar}} {\sqrt{2m[E-V(x)]}} = 
{1\over{\hbar}} {\sqrt{-2m{\mu}(x-a)}} }
and
\eqqn\jnfd{
\int_x^ a k(x')\, dx' = {2 \over 3}(-z)^{3/2} \quad . }
Finally, since $\sin{\left(\phi +{{\pi} \over 4}\right)} =
\cos{\left({\phi} -{{\pi} \over 4}\right)}$ we can derived the following
matching expression
\blankline
\eqqn\hdsw{
{2\over{\sqrt{k(x)}}} {\cos {\left(
{{\int_x^ a k(x')\, dx'} -
{{\pi}\over 4}} \right)}} {\longleftrightarrow}
{{1\over{\sqrt{\beta(x)}}}{\exp {\left(
-{\int_a^ x \beta(x')\, dx'}\right)}}} \quad . }
\blankline
Since $\cos {\left({\phi} + {{\pi} \over 4}\right)}=
-\sin{\left({\phi} - {{\pi}\over 4} \right)}$ we find
\blankline
\eqqn\hlpt{
{1\over{\sqrt {k(x)}}}{\sin{\left({{\int_x^ a k(x')\, dx'}
-{{\pi}\over4}}\right)}} {\longleftrightarrow}
-{{1\over{\sqrt{\beta(x)}}}{\exp{\left({\int_a^ x \beta(x')\, dx'}
\right)}}} \quad . }
The connection relations for the case of a decreasing potential
$({dV}/{dx}<0)$ are given analogously
\blankline
\eqqn\evvw{
{1\over{\sqrt {\beta(x)}}} {\exp{\left({-\int_x^ a \beta(x')\, dx'}
\right)}} {\longleftrightarrow} {2\over{\sqrt {k(x)}}} {\cos
{\left({{\int_a^ x k(x')\, dx'}-{{\pi}\over4}}\right)}} }
and
\eqqn\wevv{
-{1\over{\sqrt {\beta(x)}}} {\exp{\left({\int_x^ a \beta(x')\, dx'}
\right)}} {\longleftrightarrow}
{1\over{\sqrt {k(x)}}} {\sin
{\left({{\int_a^ x k(x')\, dx'}-{{\pi}\over4}}\right)}} \quad . }
\blankline
With these connection formulae, it is straightforward to determine
the transmission coefficient T with the knowledge of the semiclassical
wave function 
\eqqn\azem{
\psi(x) = 
{1\over{\sqrt{ k(x)}}} \left( {C_{+}} e^{i\sigma_*} + {C_{-}} e^{-i\sigma_*} 
\right) }
and its asymptotic limit
\eqqn\sbea{
\psi(x) \sim \cases{ \psi_{Iin}(x)     & for $ x\ll a$ \cr
                     \psi_{IIIout}(x)  & for $ b\ll x$ \cr}  }

\noindent
in the region to the far 
left and right of the barrier, which
are given by
\eqqn\abla{
\eqalign{
\psi_{Iin}(x) =& - {B \over {\sqrt{ k(x) }}}
                      [ \exp \left( - \int_a^b \beta (x) dx\right) \sin 
\left(\int_x^a k(x^\prime) dx^\prime -{\pi\over 4} \right) +\cr
+& 4 i \exp \left( \int_a^b \beta (x) dx\right) \sin\left(
\int_x^a k(x^\prime) dx^\prime + {\pi\over 4} \right) ] \cr} }
and
\eqqn\bbl{
\psi_{IIIout}(x) = {{2B} \over {\sqrt{k(x)}}}
\exp \left( i\int_b^x k(x^\prime) dx^\prime -i{\pi\over 4} \right) }
let us make an immediate determination of the 
transmission coefficient T.
This probability of tunneling is defined by means of the ratio between
the probability current density
\eqqn\abr{
j=Re\left({{\hbar}\over{im}}{\psi}^* {{d{\psi}}\over{dx}}\right)=
-i{{\hbar}\over{2m}}\left({\psi}^*
{{d{\psi}}\over{dx}}-{\psi}{{d{\psi}^*}\over{dx}}\right) }

\noindent
of the transmitted and the incident waves: 
\eqqn\acf{
T= { {j_{III}} \over{j_I} } }
 
\noindent
where
\eqqn\vbe{
\cases{ j_{III} = 4 {\vert B \vert}^2 {\hbar \over m}  \cr
  \cr
j_I = 4 {\vert B \vert}^2 \left[ \left( e^{\sigma_*} + {1\over 4} e^{-\sigma_*} 
\right)^2  - \left( e^{\sigma_*} - {1\over 4} e^{-\sigma_*} 
\right)^2 \right] {\hbar \over m} \cr } }

\noindent
with 
\eqqn\ult{
\sigma_*=\int_a^ b \beta(x)\,
dx={{1}\over{\hbar}}{\int_a^ b}\sqrt{2m[V(x)-E]}\, dx\quad .}

\noindent
Thus the WKB techniques show that in the semiclassical limit, the 
probability of tunneling is given by 
\eqqn\uult{
T = {{j_{III}}\over{j_I}}={{e^ {-2{{\sigma}_*}}}\over
{{\left(1+{{1}\over{4}}e^ {-2{{\sigma}_*}}\right)}^ 2}}
\simeq   e^ {-2{{\sigma}_*}} }

\noindent
that is valid under the assumption that $ \exp ({- 2\sigma_*})  \ll 1$.
Therefore the probability of being
reflected is correspondingly reduced to $R=1-T.$
\blankline
\leftline {\bf III. Effective Semiclassical Approximation
to Barrier Penetration}
\blankline\noindent
In the complex method we previously analyzed, no attempt is made to 
clarify the contributions of the multipliers $C_{\pm}(x)$ which
indeed are $x$-dependent and 
associated to the modifying effects of the possibility of internal
reflections inside the barrier, in the sense that the particle can
travel from turning point $a$ to turning $b$ in a several $n$ number of
ways. A systematic account of multiply reflected contributions in the
limit of a continuous potential is associated with the following coupled 
first-order equations
\eqqn\ufgrt{ 
C'_{\pm}(x)={{\sigma}'}_{1}(x)C_{\mp}(x)\exp\left(\mp {2i\over{\hbar}}
w(x_0,x)\right)
}
where the prime refers to the derivation with respect to $x$
\ref\FF{ N. Fr\"oman and P. O. Fr\"oman,
''{\it JWKB Approximation 
}'' (Amsterdam, North-Holland, 1965)}.
These 
coupled equations are formally equivalent to the Schr\"odinger's Eq.(1).
The different solutions of the Schr\"odinger's equation are obtained by 
applying different conditions to $C_{\pm}(x)$.
In particular, imposing that there is no reflected wave far beyond
the potential barrier and assuming that the incident wave has unit 
intensity, we obtain the scattering solution, after a trivial 
change of normalization, 
\eqqn\solut{
\cases{
{C_+}(x)=1+{\int_{-\infty}^ x dx'\, r(x'){C_-}(x')
\exp{\left[{-{2i\over{\hbar}}w(x_0,x')}\right]}}\cr
 \cr
{C_-}(x)=-{\int_x^ {+\infty} dx'\, r(x'){C_+}(x')
\exp{\left[{+{2i\over{\hbar}}w(x_0,x')}\right]}} \quad .\cr}}
Wherever the differential reflection coefficient
\eqqn\rcoef{
r(x)=-{{\sigma}'_1}(x)={ {p'(x)} \over {2p(x)} }}
may be set equal to zero, a constant value for the ${C_{\pm}}(x)$
are obtained, as we mentioned above. Therefore, the precise value
of the reflection coefficient may be found with the help of successive
integration by parts
\blankline
\eqqn\parts{
\eqalign{
R= & -\int_{-\infty}^ {+\infty} dx \, r(x) C_+(x) \exp\left[+
{2i \over \hbar}w(x_0,x)\right] = -\int_{-\infty}^ {+\infty} dx \,
r(x) \exp\left[+{2i \over \hbar}w(x_0,x)\right]-\cr
- & \int_{-\infty}^ {+\infty} dx \, r(x) 
\left\{\int_{-\infty}^ x dx' \, r(x') C_-(x') \exp \left[-
{2i \over \hbar}w(x_0,x')\right]\right\}\exp \left[{2i \over \hbar}
w(x_0,x)\right] \quad . \cr}}
\blankline
In this case, however, the main contributions to the integral
cannot be selected easily because they depend on the 
analytic properties of the function $p(x)$ and, in the last view, on the
type of the singularities of the potential. In fact, the Schr\"odinger 
Eq.(1) takes the form 
\eqqn\screq{
{ { {d^2}\phi(w) }\over{d{w^2}}}+{\left[{1\over{{\hbar}^2}}+
{\widetilde V}(w)\right]}
{\phi(w)}=0}
\blankline
where we change the variable to the phase $w$ of the exponent in 
Eq.(54)
and make the transformation 
\eqqn\tans{
\phi(x)={\sqrt{p(x)\over{\hbar}}}\psi(x)}
being $\widetilde V$ determined by $p$ through the relation
\vskip 6 pt\noindent
\eqqn\rela{
{\widetilde V}={{3{(p')}^2-2pp''}\over{4p^4}}=
{{{{\sigma}_1}''+{{{\sigma}_1}'}^2}\over{{{{\sigma}_0}'}^2}}=
-{2\over{p(x)}}{{\sigma}_2}' \quad .}
\vskip 6 pt\noindent
Here prime denotes differentiation with respect to $x$. 
Nevertheless, we may stress that the most important quantity results the phase
$w$ of the exponent. The subtleties involved in the evaluation of the
precise value of the reflection coefficient stem on the convergent
expansion of the total wave amplitude which can be given
in momentum space as
\blankline
\eqqn\ceas{
\eqalign{
{\widetilde {\psi}}(k_0,k)= &{\,} {\widetilde {\phi}}(k_0,k)+
\int_{}^ {} dk' \,{\widetilde {G_0}}(k_0,k'){\widetilde V}(k_0,k')
{\widetilde {\phi}}(k_0,k')+\cr
+& \int_{}^ {} dk^{\prime \prime} \, \left[{\int_{}^ {}dk' \, 
{\widetilde {G_0}}(k_0,k){\widetilde V}(k_0,k'){\widetilde {G_0}}
(k',k^{\prime \prime}){\widetilde V}(k',k^{\prime \prime})
{\widetilde {\phi}}(k',k^{\prime \prime})}\right]+...
\cr}
}
\blankline
where ${\widetilde {\phi}}$ represents the free wave solution and
\eqqn\repr{
{{\widetilde G}_0}(k)={{{\hbar}^2}\over{2{\pi}}}\,
\left[{1 \over {1-({\hbar}k)^2}}\right]}
is its relative propagator. Both $\phi _0$ and ${\widetilde {G_0}}$
are obtained assuming that the 
perturbation ${\widetilde V}$ vanishes in Eq.(58). Using these
asymptotic solutions, we derive that the exact solution of reflection
coefficient is perturbatively given by
\blankline
\eqqn\given{
\eqalign{
R={\widetilde v}(k_i,k_f)+ & \int_{}^ {} dk_1 \,
{\widetilde v}(k_i,k_1) {\widetilde G_0} (k_1) 
{\widetilde v}(k_1,k_f) + \cr
+ & \int \! \! \int dk_1 dk_2 \,[
{\widetilde v}(k_i, k_1) {\widetilde G_0}(k_1)
{\widetilde v}(k_1, k_2) {\widetilde G_0} (k_2)
{\widetilde v}(k_2,k_f) ]+... \cr}}
where ${\widetilde v}(k, k')$
are the matrix elements of the perturbation ${\widetilde V}$ in
Eq.(58)
\eqqn\til{
\eqalign{
{\widetilde v}(k,k')= & \int d\xi {\widetilde \phi}^*(k)
{\widetilde V}(w(\xi))
{\widetilde \phi}(k^ \prime) \cr
= & \int_{-{\infty}}^ {+{\infty}} e^{{i \over {\hbar}}(k'-k){\xi}}
{\widetilde V}(w(\xi)) \, d{\xi} \quad. \cr}}
Such results would need to take account of all the singularities
of the perturbation ${\widetilde V}$ and indeed of $p(x)$. Thus
any value of $R$ will not in general be single valued because of
the branch points at the real turning points $x_c^ {(i)}$ where
$p(x_c^ {(i)})=0$, unless we adopt the convention of dividing the
complex Gauss plane into two Riemann sheets as shown in Fig.(1).
\blankline
\leftline {\bf Concluding Remarks}
\blankline\noindent
The propagator (Green's function) technique in the solution of 
problems in non relativistic Quantum Mechanics becomes relevant
in calculating explicitly the reflection coefficient in the case
of barrier scattering represented by an analytic function $V(x)$.
In the case of ordinary barrier penetration $E<V_{max}$, $k(x)$,
defined in Eq.(9), is in general two-sheeted, with two branch 
turning points $x_c^ {(1)}$, $x_c^ {(2)}$ located on the real axis
and we may choose the defining branch cut to connect them (Fig. 1).
An alternative approach to the semiclassical approximation allows a very
appealing picture of the transmission coefficient and generates
results identical to the usual WKB prescription, but without the
cumbersome connection formulas.
This 
method consists in the analytic continuation into the complex momentum
variable and contour integrations, wherein it is permitted. Such a 
method of complex variable in modern theoretical physics is extensively
adopted to clarify the concepts of analyticity in S-matrix theory. It
results particularly suitable for discussing the problem of one 
dimensional barrier penetration if we reconsider that the generic
propagation from $x_1$ to $x_2$ far to the left and right of the barrier
respectively in opposite sides can be considered as occurring in  
three successive steps according to the decomposition of the outstanding
integral
\blankline
\eqqn\fwok{
\left(\int_{x_{in}}^ {x_{fin}} k(x)\, dx \right) =
\left({\int_{x_{in}}^ a k(x)\, dx} + {\int_a^ b {i\beta(x)}\, dx} +
{\int_b^ {x_{fin}} k(x)\, dx}\right) \quad . }
\blankline
The complexification of the problem generates a more intuitive picture
of the physics of the process and offers an alternative technique to avoid
singularities which occur in the standard WKB.
The propagator can be, then, expressed in terms of contour integrals
which connects the initial and final points $x_{in}$, $x_{fin}$,
both of which are located on the real axis far to the left of the
barrier.
Furthermore we assert that if the incident
wave is initially located in $x_{in}$ far to
the left of the barrier, then the reflected wave is given by the 
analytic continuation of the functions involved, evaluated on the 
other side of the cut. 
It is important to note that they are on different sheets so that
any independent contour of integration has to pass through
the cut.
Of course, the singularities of the function
${\widetilde V}$ in the complex plane may involve other branch
points, but we may discard them here. Anyway, the singularities of
${\widetilde V}$ are related to the singularities of the function
$p(x)$. Clearly, the multiple integrals resulting in Eq.(63)
correspond to the effect of multiple reflections.
The contributions of the only once--reflected waves to the reflection 
coefficient are given by
\blankline
\eqqn\mult{
R \simeq -\int_{-\infty}^ {+\infty} dx \, r(x) e^{2iw(x_0,x)/ \hbar}
\quad .}
Actually, one could include also contributions from contours which 
loop the branch cut several times. These higher order terms are, in 
general, unreliable although they are expected exponentially much
smaller than the polynomial corrections which lie outside the usual
semiclassical WKB approximation.  
\blankline
\centerline {\bf Acknowledgments}
\blankline\noindent
We are grateful to the warm hospitality of the 
Department of Mathematics, University of 
Basilicata, Potenza. One of us (C.D.)  wishes to thank Guido Lupacciolu 
for his interest in the early stages of this work.
His critical comments, we profited, unfortunately, missed to be exerted
into this final version, for his untimely decease.
\vfill\eject\immediate\closeout\rfile
\centerline{{\bf References}}\bigskip
\input refs.tmp\vfill\eject
 \bye